# Toward non-textual representation of social anthropology

## Modeling cultures as knowledge graphs

**Manolis Peponakis, Sarantos Kapidakis, Martin Doerr and Eirini Tountasaki**

This version of the article has been accepted for publication in the *Journal of Digital Social Research*.

### Abstract

The study examines the emerging field of knowledge representation in the context of the semantic web and linked data, with a focus on knowledge produced within the social sciences - particularly in sociocultural anthropology. It starts from the premise that natural language, especially its textualized form, has long been the primary vehicle for producing and communicating anthropological research. Informed by theoretical approaches from information science, the study explores how computational methods may offer alternative modes of structuring and representing anthropological knowledge. It challenges the dominance of text as the sole representational medium and highlights the potential of semantic modeling to open new epistemological pathways. At the same time, it acknowledges the conceptual and methodological challenges involved in such a transition. This approach shifts emphasis away from metrics and programming, foregrounding processes of conceptualization, semantics, meaning, and reasoning as key to engaging with anthropological knowledge in digital environments.

Keywords: computational social anthropology; knowledge representation; semantic web; ontologies; knowledge graphs; computational ethnography

## 1. Introduction

Natural language and, more precisely, its textualization, has been the main carrier for the production and representation of knowledge in social sciences, and particularly in anthropological research. Despite the dominance of text, there have been steps toward using computational methods and digital-born data for anthropological research. To name a few such cases, there is a recorded relation between big data and ethnographic data (Bjerre-Nielsen and Glavind 2022; Curran 2013), as well as the contribution of machine learning to predicting human reactions (Munk, Jacomy, and Knudsen 2022). Yet, irrespective of the original means used for this knowledge generation, meaning either traditional or computational methods, there have been no relevant steps toward representing the overall generated knowledge in a machine-readable way. Because it is not only computational methods which support knowledge production, there are also computational methods for knowledge representation. These methods have their roots in the domain of information science and create a novel potential, more suitable for algorithmic processing, for recording and representing knowledge in diverse domains. Following these approaches, the development of artificial languages is considered essential for representing knowledge.

Geertz mentions that "The essential vocation of interpretive anthropology is not to answer our deepest questions, but to make available to us answers that others, guarding other sheep in other valleys, have given, and thus to include them in the consultable record of what man has said" (Geertz 1973, 30). And Marcus wonders whether all this knowledge generated by anthropologists can add up in the form of a cumulative archive (Marcus 1998). The question that arises is how this archive -holding what humanity has said- could be structured, so that it can be highly findable, accessible and reusable. Is natural language alone the most suitable mechanism for achieving this, or could artificial languages help in this regard?

In this context, the study puts into the spotlight some emerging methodological approaches for representing knowledge generated through qualitative analysis in the domain of social sciences by means that are effective for computational processing. For this purpose, emphasis is given to -and examples are drawn from- the sub-domain of social anthropology, deploying the potential arising from modeling knowledge representation for the semantic web. Seaver, an anthropologist who claims that algorithms are inherently cultural, notes that we are not distant observers but active enactors who produce 'algorithms' through our research (Seaver 2017, 5). The representation of these 'algorithms' in a 'language' suitable for efficient computational processing directly affects the potential of disseminating and processing anthropology in the digital infrastructures.

Studies have shown that there is an ongoing transition from the instrumentalization of computers as tools for calculations to the implementation of computer and information science theory and methodologies for analysis, deduction, knowledge representation, and reasoning, as part of the research process (Peponakis et al. 2024). Building on this evidence, the current study does not suggest a violent and radical methodological passage from text-based anthropology to non-textual representations; nor does it suggest a global methodological transition to quantifications and metrics. Instead, it introduces new conceptualizations for representing qualitative analysis in social sciences, setting new horizons in terms of defining semantics and meaning. The ultimate goal is to lead to effective reasoning by using computational methods.

## 2. Aims and objectives

The ultimate aim of the study is to contribute to new conceptualizations regarding the methods used for representing qualitative analysis in social sciences. Therefore, the main goals are, first, to explore the dynamics and the potential of representing qualitative analysis in the social sciences in the context of the semantic web, and, second, to highlight the benefits of this approach as well as the challenges that arise for social scientists. For these purposes, the study refers specifically to the domain of social anthropology, which is a domain largely based on textual representation for delivering research results. The specific objectives for achieving the study's main goals are as follows:

- highlight the dominant role of natural language as a form of representation for anthropological research
- introduce certain concept-centric rather than scriptocentric artificial languages used for knowledge representation, and
- introduce fundamental principles for representing knowledge in the semantic web in general, while also focusing on the representation of anthropological knowledge, and outline the requirements and challenges for representing social anthropology in the semantic web.

This study is not about ethnographic fieldwork in the digital environment, as in the studies of Hine (2000), Boellstorff (2015), and Forberg (2021), to name a few. Instead, it is about representing data and the analysis of those data in the context of the semantic web, regardless of whether the fieldwork took place in the digital environment.

Finally, it should be clarified that, in anthropological research, there is a distinction between the recording of ethnographic data and their subsequent representation and interpretation. The former refers to the systematic collection and documentation of data -such as field notes, audiovisual recordings, or sensor-based traces of human activity- which constitute the empirical foundation of inquiry. However, anthropology engages in this process primarily serves as a necessary precondition for the latter: the interpretive work through which data are contextualized, analyzed, and rendered meaningful within broader theoretical frameworks. Accordingly, the present study is not concerned with methodological or technical issues related to the capture of primary data per se. Questions surrounding the production of raw ethnographic records -whether through observation, recording devices, or computational sensing technologies- fall outside its scope.

## 3. Challenges in representing knowledge

Natural language -as a means or carrier for representing knowledge- is compared against artificial formal languages used for these purposes. Even before entering this discussion, we acknowledge the existence of different views about whether it is even possible to represent knowledge in any way. For example, there are viewpoints arguing that knowledge is only present in the human brain; so, anything represented in the form of texts or images is simply information. Yet, in the context of this research, the term "knowledge" is used to refer to a type of sophisticated information representation, mainly because we do not just refer to the process of recording any information, but to a higher level of abstraction. Besides, the term is widely accepted and adequately defined in the context of the DIKW hierarchy or pyramid, where DIKW stands for Data-Information-Knowledge-Wisdom (Frické 2019). Following a bottom-up approach of the pyramid, the identified hierarchical levels are: "data", which is used for relatively unstructured forms of statements or facts; "information", which is used for data that have been processed, further organized, and, perhaps assigned a semantic determinant; "knowledge", which is used when information can drive decisions; and finally, at the upper level of the hierarchy, there is "wisdom". This last characteristic brings out the human element in this approach, addressing the discussion mentioned above, regarding whether knowledge is the semantic difference between humans and computers. The level of "wisdom" in this hierarchy defines the boundary as to what the product of computational processing can be, and the circumstances under which the human contribution is irreplaceable.

### *3.1 Natural language as a means for representing knowledge*

Words in natural language represent concepts; therefore, they are essential for thinking and communicating. But the relation between words and concepts is not a "one-to-one" relationship, leading to the phenomenon of ambiguity. For example, a word may refer to multiple concepts (polysemy) or a concept may be expressed by different words (synonymy). This is why natural language and context are critically interrelated factors. Context is not only related to language but also to a wider cultural framework, since an implicit background of cultural information is involved during verbal communication. Pinker (2007, 227) highlights the contribution of context by considering "how much knowledge of human behavior must be interpolated to understand what *he* means in a simple dialogue like this:

Woman: I'm leaving you.
Man: Who is he?"

Still, many times, this implicit knowledge, which is inherent in humans, does not allow the signal to be transmitted the way its creator intended. And it is not about a potentially simple misunderstanding, but about the subjective hermeneutics which have developed particularly in the post-modern era, mainly through literary criticism, particularly following the work of Jacques Derrida. Post-modernism claims that objective communication is impossible since the receiver of the message could not know with certainty what the intended meaning of the message was.

Abrams goes a bit further with this approach and critiques it by acknowledging that we cannot guarantee that any interpretation is certain and true by reference to an absolute and eternal criterion beyond the regularities of our shared linguistic practice. Therefore, he explains, our certainty about an interpretation and its truth is never unconditional, since there is always the possibility, although in some cases relatively distant, of making a mistake. In this way, he undermines the extravagant position of theorists of this movement, noting that according to their own theory even the theorists themselves do not exist in their alleged theoretical world. And he illustrates his position by noting that, in everyday situations and interactions, the post-modernists cannot but abandon their theoretical language and adopt human language. This language refers to persons who have underlying intentions when using certain words, who want to communicate certain meanings, and who are held intellectually and morally accountable for their statements (Abrams 2014), meaning that they are imbued with intent and purpose.

Additionally, we must consider the ongoing transformation of natural language, as well as that of concepts, whether through the addition of new words or shifts in the meanings of existing ones; hence, language is considered a *living entity*. This attribute of language makes Feyerabend notes that, without a constant misuse of language, there cannot be any discovery, any progress (Feyerabend 2009, 18). At this point, we cannot but acknowledge that this paradox, namely the assignment of a new concept to a known word which was previously used with different semantics, is mainly an inherent function performed by the human brain. For example, the word *book* is now used to describe not only printed material, meaning what we have accepted as being the traditional format of books, but also digital files. Another example of our brain assigning a previously defined symbol to a new use is the widely recognized email icon, which depicts an envelope. Similarly, smartphones use the icon of a traditional landline telephone handset -commonly used in the 1980s- to represent the function of making phone calls.

Overall, formalistic languages may provide us with a solution for dealing with ambiguity, which is an inherent attribute of language and, therefore, cannot be overlooked. On this matter, an expert on controlled natural languages notes that "Natural language is very imprecise in this sense, because a large amount of context information is needed to grasp the meaning of typical sentences. Formal logic languages, on the other hand, have maximal precision, because their meaning is strictly defined solely based on the possible sequences of their language symbols" (T. Kuhn 2014, 128). In this way, it becomes apparent that the use of a concept-centric formalistic language is essential for representing context, because concepts [and not words] are the building blocks of thoughts, which makes them crucial for any psychological process of categorization, inference, memory, learning and decision making (Margolis and Laurence 2011, para. 1). And, as cognitive scientists propose, concepts are mental representations which refer to categories (Rips, Smith, and Medin 2013, 177).

#### *3.1.1 Natural language, script, and anthropological knowledge*

A fundamental problem of social and human sciences is that their descriptions of social phenomena are strictly related to specific natural languages (Van Der Leeuw 2004, 123). Social anthropology is no exception. The etymology of the word *anthropology* derives from the combination of two Greek words, namely *ánthrōpos*, which means human, and *-logíā*, which is a suffix denoting the study of something or that which is said; the etymology of the suffix derives from the verb *légō*, which means *speak*, *tell*, or *say*. Correspondingly, the etymology of the word ethnography derives from the Greek words *éthnos*, which means nation, and the suffix *-graphia*, which is, again, a suffix denoting the study of something or, in this case, what is written, since the etymology of the suffix derives from the Greek verb *gráphō*, meaning *write*. Etymology may show the roots of a term but does not define the course of the scientific domain to which it refers. For example, the word biology also refers to discourse but its research and representation are not exclusively scriptocentric. Not only did anthropologists refrain from interrogating the hegemonic role of natural language in the representation of knowledge, but they actively reinforced its privileged status. Mueller emphasizes that ethnography, both as a process and as a product, is mainly verbal and scriptocentric (Mueller 2016, 99) and Starn notes "By now, writing is unavoidably part of who I am -in fact, I don't know how to do much else. My grandfather was a mechanic; he fixed cars. I'm an anthropologist; I write" (Starn 2022). In this context, it is not surprising that one of the most influential anthropological texts, with thousands of references, is the book *Writing Culture* (Clifford and Marcus 1986).

Many scholars treat the aforementioned work as a reference point for discussing "the crisis of representation" (Kuper 2003). Yet, the emphasis on textual representation, clearly driven by the theories of Jacques Derrida, triggered the turn of anthropology to Humanities viewed from the perspective of post-modern theory, while distancing itself from the social sciences. In the afterword of *Writing Culture*, which is authored by Marcus himself, it is emphasized that textualization is at the heart of ethnographic endeavours, both in the field and in the academic ecosystem (Clifford and Marcus 1986, 264). In the same work, James Clifford expresses, first, the opinion that ethnographic practice effectively textualizes the other (Clifford and Marcus 1986, 111) and repeats later that ethnographic practice is mostly a process of writing and of textualization. Almost twenty years after this publication, Marcus evaluates its influence and finds that no practical shift from textual representation is observed in the domain of ethnography (Marcus 2007). Another cumulative work on the history of textualization is *Modernist anthropology: From Fieldwork to Text* (Manganaro 1990), where a whole chapter is devoted to *Ethnography as Discourse*, which examines how ethnographic writing constructs cultural meaning through narrative strategies.

Presumably, it is not natural language overall, but its textualization that becomes the essential means of producing and representing anthropological knowledge. In the following section, this approach is critically examined to determine whether it is the unique method available.

#### *3.1.2 Expressive power: script as an example of natural language representation; information science as a means for information representation*

In a world without writing (script), there would be no basic mechanism for producing texts either. In such a world and at some point, there would probably be an effort to introduce a scripting system, perhaps in a way that current alphabets in the western world did. Focusing on the weaknesses of such an effort, and considering what the sounds of natural language provide in the communication process, an argument arises concerning the expressive power of this system. In terms of representing language overall, it is not adequately sophisticated. First, it fails to represent two basic characteristics of verbal communication: the volume and vocal intensity. In addition, it does not have a mechanism for representing the resonance, the accent, or the tone of voice, and many more characteristics of sounds which are detectable and processed by human hearing. On the other hand, it uses few, overly simplistic symbols, such as the exclamation mark and the question mark, which could not even represent the slightest ironic intent in a sentence. Moreover, there is no way of representing pauses between words nor is there a way of representing the duration of syllables, etc[1]. If the new system is less expressive than the existing one, i.e. oral language, then the question is: *Why does this new system even exist?*

Galeano's critique of script seems to fit within the context of the aforementioned arguments. In his work "*Writing No*" he notes: "Some five thousand years before Champollion, the god Thoth traveled to Thebes and offered King Thamus of Egypt the art of writing. He explained hieroglyphs and said that writing was the best remedy for poor memory and feeble knowledge. The king refused the gift: "Memory? Knowledge? This invention will encourage forgetting. Knowledge resides in truth, not in its appearance. One cannot remember with the memory of another. Men will record, but they won't recall. They will repeat, but they will not live. They will learn of many things, but they won't understand a thing" (Galeano 2009).

[1] This limited expressiveness of script drove modern linguistics, and especially the sub-domain of phonology, to develop specified "alphabets" which allow better recording of phonemes.

Leaving behind the hypothetical universe and mythology, in the modern western world we have experienced a massive downgrade of this argument. Since the Renaissance, the role of script has been so fundamental for peoples in the West that Foucault mentions "Henceforth, it is the primal nature of language to be written. The sounds made by voices provide no more than a transitory and precarious translation of it" (Foucault 2007, 42). An analogy can be viewed in the case of computer science. Representation and data management using computers set restrictions for what is transferred through voice or through what is written on paper; but a new language is formed and a new world is introduced for representing and managing information which could bring more dramatic changes than those introduced by the invention of script.

Another analogy between script and computer science lies in the fact that there is no single alphabet representing every natural language: it has to do with the existence of different requirements in order to respond to diverse needs. Similarly, the development of modern information systems considers pre-existing specifications and conditions, and makes sure that they are efficiently managed by the new systems. Building on our discussion of script, we realize that it encompasses more than just natural language. In some cases, natural language was proven to be limited in terms of formalism, so that certain findings or hypotheses of a domain could not be represented. Then, these domains developed their own languages, like mathematics and programming or markup languages, which maintain their own script and syntax. Along the same lines, specific scientific domains must develop their own "scripts", so that they can extensively exploit the potential of computer science. In this case, the situation tends to be more complicated because the subjects are algorithms, knowledge representations etc.

When specific domains realize the fundamental principles of modeling as required in the context of computer and information science, and they begin not just to use it but also to expand it to meet their own needs, the growth of research and its results may accelerate. From this viewpoint, the main concern would not be about using software for text processing, like the one used for drafting this article, or about using another type of software for quantified analysis of datasets. Instead, the focus would be on how data can be codified for computational processing. The domain of knowledge representation is a fertile ground that can help the social sciences achieve this goal.

### *3.2 Artificial languages and knowledge representation: A non scriptocentric approach*

Knowledge representation, often studied alongside reasoning, is a domain of artificial intelligence (AI) that focuses on encoding information so it can be used in computers to perform complex functions and solve problems. For knowledge representation, three main components are needed: syntax, semantics, and reasoning. Syntax allows the construction of well-formed statements, following the same rationale as natural language syntax. Semantics defines the meaning, i.e. what is intended to be represented. And finally, reasoning is the analysis process aiming to extract conclusions derived from the data in question. The above components establish a generic framework that allows the development of artificial languages, which can be applied following more specific rules.

The most essential element in knowledge representation is meaning. As Guarino notes, the aim is to limit the number of alternative hermeneutics by defining the meaning of basic ontological categories which are used to describe a specific domain; thus, the ontological level is the level of meaning (Guarino 1995, 632). This section presents the basic principles for codifying meaning in the contemporary digital environment.

#### *3.2.1 The semantic web as a knowledge graph*

At this point it is essential to underline that the relationship between artificial intelligence and machine learning is often presented as one-dimensional. Machine learning is one aspect of artificial intelligence; but the domain of knowledge representation deals with codifying existing knowledge in a way that the machine "understands", so that it can perform reasoning, through which new knowledge is formed. It should be clear that machine learning is a process that happens within the machine, forming "non-human understanding" with regard to the outcome of reasoning; while knowledge representation has to do with defining the way in which human thinking can be a building block of the machine. Antoniou and Van Harmelen note: there are two main solutions for achieving artificial intelligence. The first solution is the development of very sophisticated techniques for data analysis. The alternative solution focuses on data formats and is oriented more towards data structures so that data processing is facilitated; in this approach, structure enables sophisticated processing. The latter approach is the one paving the way towards the semantic web (Antoniou and Van Harmelen 2008, 3).

Semantic web or the web of data or web 3.0[2] (not to be confused with Web3) is an extension of the current web and focuses on the structure of data. The aim is to turn to data structures that will be more effectively processable by computers, in contrast to the functionality of the contemporary scriptocentric web. Many authors treat the terms semantic web, web of data, and linked data as synonyms. Yet, Machado et al devote a study on the differences between these terms. They summarize that the aim of turning the web into a global database led to the concept of the web of data. The specifications developed or related to this aim as well as the processes and rules for achieving it are known as linked data, which make possible the publication of data in a way that allows their connection with other data from different sources and formats (Machado, Souza, and Simões 2019, 712).

Linked data is the method for connecting structured data, and they are a priceless tool for developing shared and structured information; not simply data. Knowledge representation and semantic technologies supply the mechanism for upgrading linked data to meaningful statements since they carry the intended meaning. The meaning is essential for the correct apprehension of these statements and their interconnections, in relation to the description, context, and provenance (Oldman, Doerr, and Gradmann 2016, 267).

The main language for codifying data for the semantic web is RDF (Resource Description Framework). The fundamental syntax mechanism of this language is based on statements each of which has the form of a triple. The three elements of each triple are *subject*, *predicate* and *object* (W3C 2014). This structure is like the structure of natural language, namely *subject*-*verb*-*object*. The basic principle is that there are nodes and edges, both of which are assigned to URIs[3]. An analogy for natural language could be that the nodes are the nouns, while the edges are the verbs. In graphical representations the nodes are the elliptical shapes, while the edges -i.e, the connecting lines- are the predicates. The rectangles refer to textual values which cannot be the subject for defining new relations, meaning that the values are the names of the entities.

[2] The term Web 1.0 refers to the first phase of the Web's development, during which interaction between end users and available content was extremely limited. Web 2.0 refers to the subsequent phase of the Web, in which end users could easily create, share, and publish content. This was made possible largely by the development of social media technologies and platforms such as YouTube and Facebook.

[3] URI is the acronym of Uniform Resource Identifier, which is a string of characters identifying unambiguously a resource. For uniformity purposes, all URI types follow certain syntactic rules. One of the most popular URI forms is the one based on the http protocol where every URI stands for an entity/resource.

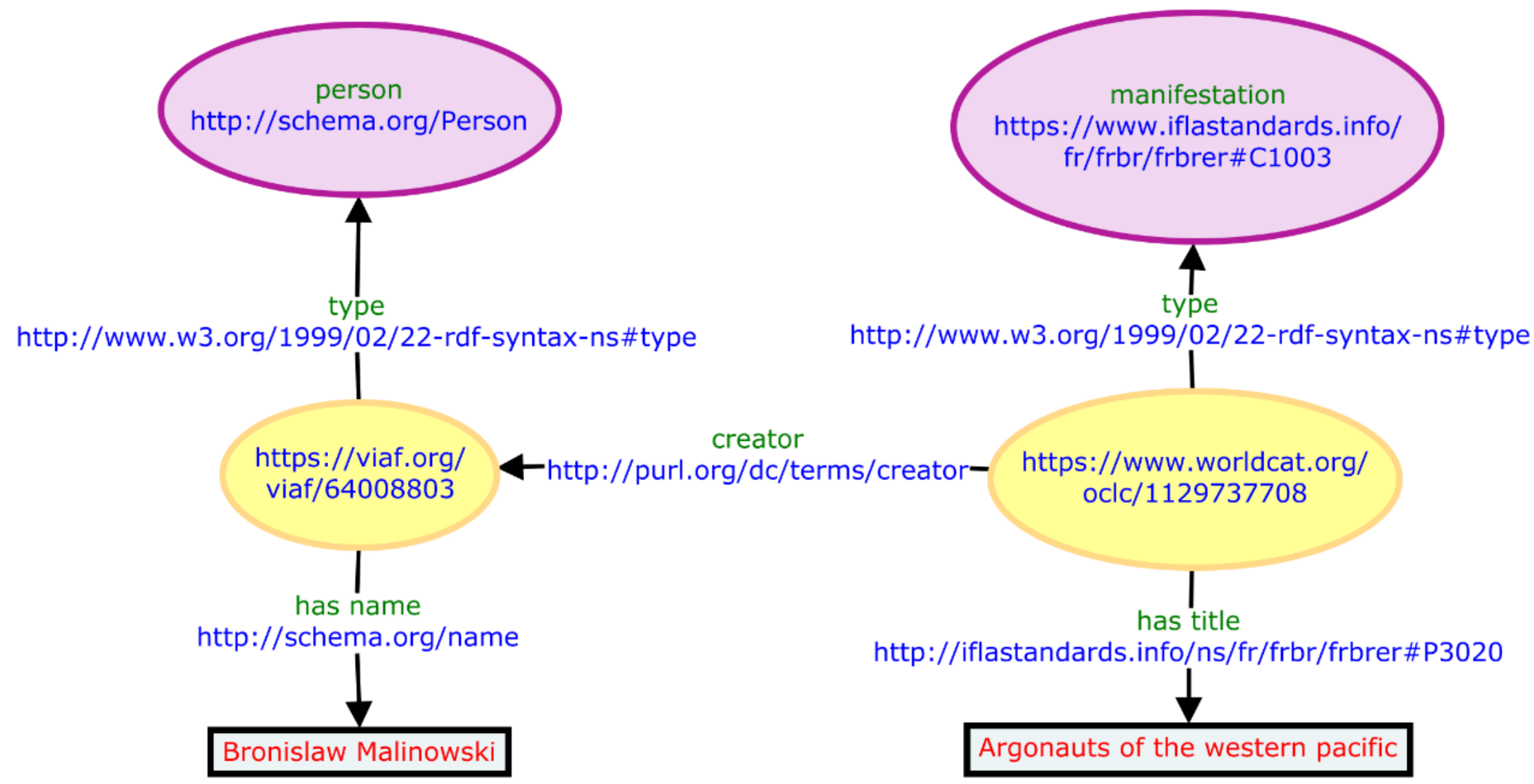

**Figure 1.** An example of RDF data structure for the semantic web (graphical representation)
*The words in green do not form part of the data codification; they are used only to aid in reading the graph.*

According to the aforementioned presentation of what the elements of RDF are, Figure 1 states that the left yellow node, http://viaf.org/viaf/64008803, is the creator of the right yellow node, http://www.worldcat.org/oclc/1129737708. But what is depicted here is not reality; the nodes function as surrogates of entities, meaning that the information actually represented is that the entity assigned to the first address, i.e. Malinowski, has created the resource assigned to the second address, i.e. the work titled *Argonauts of the Western Pacific*. Sowa notes that knowledge representation is, de jure, a surrogate. Physical objects, events or relations cannot, in fact, be stored directly in a computer. But they can be represented using symbols; and these symbols can function as substitutes for external things. By managing the internal surrogates, the software in use can simulate the external systems or perform reasoning based on them (Sowa 2009, 134). Therefore, with reference to Figure 1, there is a web address functioning as a surrogate for either a category (class of things), or an entity (an instance of a class), or even a relation between things or instances.

A textual and more eloquent representation of the triples in Figure 1 would be as follows: Malinowski, who is a person, created a resource which has the title *Argonauts of the western pacific*. What is important is that the words "person", "resource", "creator", "has title" and all words in green font, are unambiguously defined according to certain standards in relevant URIs[4]. The two nodes in yellow show the basic resources that the graph holds, i.e. the record for Malinowski (http://viaf.org/viaf/64008803) from the *Virtual International Authority File* and the record for the book *Argonauts of the western pacific* (http://www.worldcat.org/oclc/1129737708) from WorldCat, the international virtual library catalogue. The classes, i.e. the nodes in purple, declare the types of entities, not the entities.

In the above example, properties and classes were deliberately not selected from a specific schema or ontology for illustrating the description of the resources included. In this way, the flexibility allowed regarding the selection of properties is demonstrated during the creation of knowledge graphs, while the obvious absence of properties related to the core of Anthropology results from the lack of corresponding established ontologies. And it is precisely this gap that this study demonstrates, pointing out that anthropology and cultural studies must be integrated into this process.

[4] "Person" is defined in http://schema.org/Person, "resource" is defined in https://www.iflastandards.info/fr/frbr/frbrer#C1003, "creator" is defined in http://purl.org/dc/terms/creator, "has title" is defined in https://www.iflastandards.info/fr/frbr/frbrer#P3020, and http://www.w3.org/1999/02/22-rdf-syntax-ns#type (rdf:type) is defined by RDF and states that a resource is an instance of a class.

Although it may be regarded as an oversimplified approach, it could be said that RDF makes possible the representation of whatever needs to be expressed in the form of structured[5] information which is machine processable. At the same time, it is also close to what can be expressed using natural language. But the modern technological environment allows for a higher level of abstraction than this. This means that we can not only extract obvious information for specific entities, like in the case depicted in Figure 1 in which the information is rather explicit; Malinowski, who belongs to the class *People*, authored the work *Argonauts of the Western Pacific*, which belongs to the class *Books*. The purpose is to declare the rules which define the structure of information in a machine-readable form. At this upper level, it is possible to represent logical rules, and then to confirm the data against these logical rules. In Figure 1 the upper level of abstraction, which is the actual target of knowledge representation, is to provide us with the conclusion that *People write Books*. To achieve this, there is another need to make a clear distinction between the syntactical and the logical correctness of the statements. For this purpose, another example follows. Let us consider the sentences "The cow gave birth to a calf" and "The chair gave birth to a stool". Both sentences are syntactically valid. But, if the applicable rule is that *only living creatures can give birth* and *Cows* are assigned to the category of *Living creatures*, while *Chairs* are not, then the second sentence, although syntactically valid, is logically false. And this conclusion can be extracted through reasoning. The resources -in the environment of semantic web- which can codify these logical rules and facilitate reasoning are called ontologies, which are one of the important pillars of the semantic web.

The concept of ontology derives from two distinct domains; traditionally, one is rooted in philosophy, while the other, more recently introduced, derives from computer and information sciences. The first approach focuses on categorical analysis, in which the intention is to record reality. The approach to ontologies deriving from computer science also focuses on recording reality, but it has a different intention: to create computationally processable models of reality, artifacts that can be used by software. In addition, the requirement is that these models are directly interpretable for the purposes of reasoning using special software called "inference engines". The main goal is that they enhance the software with human level semantics (Poli and Obrst 2010, 1). One of the most popular definitions of an ontology, in the context of computer science, is that it is an explicit specification of a conceptualization. The term is borrowed from philosophy, where Ontology is a systematic account of Existence. For AI systems, what 'exists' is what can be represented (Gruber 1995, 908). Based on the above, it becomes obvious that anyone dealing with ontologies in this context has to move away from questions that deal with reality and turn to questions that deal with the representation of reality (Kohne 2014, 88), embracing the fact that within the "inference engines", this representation functions as a surrogate for the things that exist in the world. Ontologies can be perceived as a type of technological semantic gateway since they allow communication and coordination among various disciplines, languages, classes or database schemas. This is a basic aim for cyber-infrastructures, since it allows data and resources to flow freely across institutions of science (Ribes and Bowker 2009, 201).

An ontological representation includes four basic elements, namely a) the entities/instances (objects, concepts, persons, etc), b) the classes which they belong to, c) their properties, and d) their names. To better relate these types of elements to their representation, in Figure 1 the entities are presented as yellow nodes (elliptical shapes), the names are presented in red fonts within the rectangles, the properties are the connectors (edges) and the categories (classes of things) are the purple nodes (elliptical shapes).

The most popular language for codifying ontologies is the Web Ontology Language (OWL) (W3C 2004). Beyond the characteristics already presented in the previous paragraph, OWL divides the properties into two basic types. The first type is datatype properties, i.e. properties for which the value is a data literal, and the second type is object properties, i.e. properties for which the value is an individual. For example, the properties of the first type could be used to declare the height of a person, while the properties of the second type could be used to declare the relation between two persons. For computational processing and reasoning, any instance is defined through its properties. To further elaborate on this, the "meaning" of a concept is declared by its position within the network, which means that the concept is defined by the properties which are assigned to it, and not by its definition or its name (Peponakis et al. 2019, 448).

[5] Structured data is organized according to a predefined model -such as tables or schemas-, making it more easily processable by machines. Unstructured data, by contrast, lacks a fixed format -such as text or images- and therefore requires more complex methods to interpret and analyze.

*3.2.2 Indicative example of modeling social phenomena*

We could use kinship as an illustrative case for the implementation of social phenomena as knowledge graphs, in order to demonstrate both the applicability of this approach to anthropological data and the need for more fine-grained specification of the recording elements that anthropologists are required to produce. A clear and illustrative example is marriage. In a simplified, Western-centric perspective, marriage begins with the union of two individuals; it is therefore conceptualized as a relationship that connects two persons. If we were to express this in terms of the semantic web, then it could be represented as shown in Figure 2. This diagram illustrates a conceptual level in which categories (classes) indicate how individuals may potentially be connected to one another, while also reflecting how specific individuals are in fact connected.

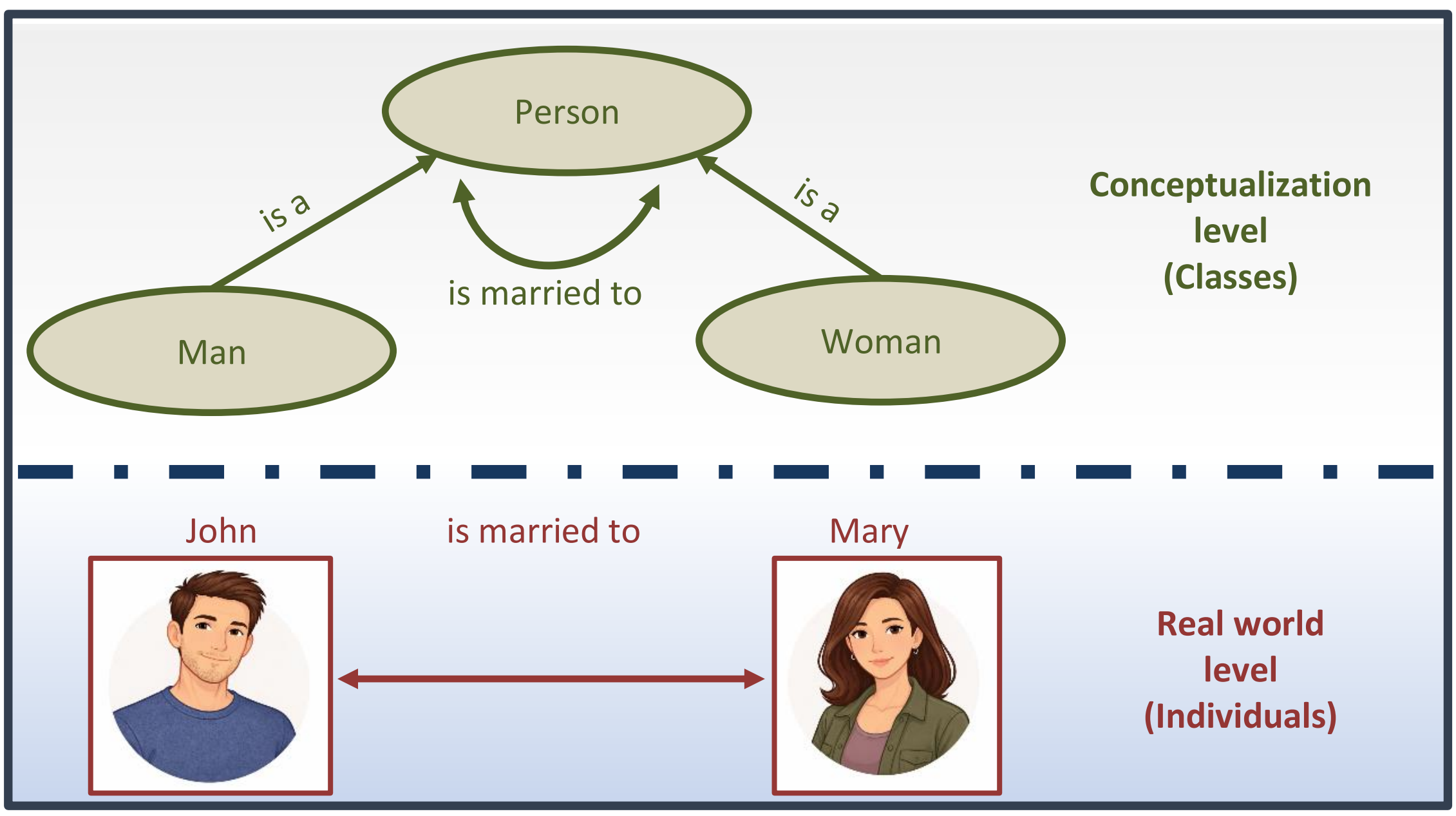

**Figure 2.** Conceptualization and real world

More specifically, Figure 2 shows, at the conceptual level, that Persons marry other Persons, and that they can be divided into the categories of Men and Women. Furthermore, the property "is married to" is symmetric (i.e., if Person A is related to Person B via this relationship, then Person B is also related to Person A via the same relationship). This is at the conceptual level. In the real world, we state that John is married to Mary. The appendix contains the relevant code in OWL. However, for the anthropological approach, the goal is not the development of the code but the development of the conceptualization level, using the logic of classes and properties, from real-world data.

How, then, could marriage be effectively modeled so that it can reflect the various forms this institution takes in every culture? Because the form of marriage depicted in the above diagram reflects a simple stereotypical model that predominated in the Western world, rather than a structure that is representative of societies globally. Consequently, questions such as which individuals may marry one another, as well as how many partners an individual may have -whether simultaneously or sequentially- constitute matters of substantive concern for anthropologists studying specific cultural contexts. The way marriage -and the kinship relations that typically arise from it- can be modeled is neither self-evident nor straightforward within the framework of the semantic web and knowledge graphs. Indeed, an entirely different modeling approach could be adopted from that illustrated above. As an example, we turn to CIDOC-CRM.

CIDOC-CRM is a highly influential ontology in the field of cultural heritage developed by the CIDOC, a committee of the International Council of Museums (ICOM). The ontology adopts an event-centric approach (Doerr 2003). Accordingly, in the case where CIDOC-CRM -or a compatible framework- were to be employed, it would be necessary to introduce a class representing marriage as an event in which individuals participate. In this type of modeling, there is no direct relationship between individuals; rather, the relationship is mediated through the event of marriage itself. This structure is evidently more expressive, as each individual marriage instance can be associated with its own set of properties. Thus, it becomes possible not only to state that an individual participates in a given marriage as a spouse, but also to specify additional attributes, such as the date of its inception, the location of the wedding ceremony, the duration of the union, and any offspring resulting from it.

At a more general level, CIDOC-CRM provides a robust mechanism for modeling events and their historical dimensions, effectively integrating temporal entities, events, actors, and the (conceptual or physical) objects they produce. In this respect, its core design demonstrates considerable potential as a candidate ontology. However, it also presents a significant limitation that restricts its applicability to a substantial portion of anthropological information: it does not accommodate the representation of entities that do not exist in the real world. As a result, it cannot adequately model deities, mythological constructs, and related phenomena. Nevertheless, its sophisticated approach to representing events and their evolution over time remains highly valuable and merits consideration, particularly for the documentation and analysis of phenomena central to anthropological research.

## 4. Knowledge representation of anthropological information in the semantic web: Advantages and main challenge

As we mentioned in the introduction, the purpose of anthropology, according to Geertz and Marcus respectively, is to make available to us the answers that others have given, and thus to include them in the consultable record of what humans have said, thereby creating a cumulative archive. The semantic web and the relevant knowledge graphs may currently represent one of the best available solutions, as they allow for the semantic integration of geo-temporal, linguistic, cultural, and relational data within a unified yet flexible framework (Chansanam et al. 2025).

Barley, stigmatizing the way in which anthropologists use their data, says that “Fact-gathering in itself has few charms. Anthropology is not short of facts but simply of anything intelligent to do with them. The notion of ‘butterfly-collection’ is familiar within the discipline and serves to characterize the endeavours of many ethnographers and failed interpreters, who simply amass neat examples of curious customs arranged by area, or alphabetically, or by evolutionary order, whatever the current style may be” (Barley 2000, 9). All the information deployed in previous sections refers to a method which allows “intelligent agents” to manage the products of anthropological knowledge effectively, so that data is better managed and added value services can be based on it.

Trying to achieve the overall goal set above, we realize the difficulty which Feldman suggests: “The absence of a coherent framework for organizing contemporary ethnographic findings reveals itself in subtle but telling ways […]. What I propose is a somewhat different mode of ethnographic theory, an approach that would draw inspiration from cross-cultural research methods and seek to avail itself of the technological advancements of our time” (Feldman 2017, 2).

### *4.1 Advantages for anthropology*

Many benefits can be attributed to the semantic web, as implemented through linked data. Some of the key advantages are: promotion of common data formats and exchange protocols on the Web which allow for the explicit definition of the intended meaning in a machine processable way; collaboration in data creation; reasoning over data; integration of heterogeneous data sources; support in data exploration, data analysis and knowledge discovery; working on multiple perspectives of the same data. In practice, the semantic web will allow us to answer questions we haven't even asked!

Obviously, what the previous paragraph mentions in general also applies to anthropology. But the representation of anthropological information in the context of the semantic web does not come with a known list of beneficial outcomes. It is an ongoing process, therefore it depends on the course of actions taken in-between, and it changes over time, depending both on technological advances and the knowledge hub created through anthropological research. This ongoing process adds more value to the overall endeavor which is difficult to predefine. Despite the challenges, this section presents some of the advantages, focusing mainly on off-the-shelf benefits.

In anthropological research, the interlinking of heterogeneous data through semantic web technologies is particularly valuable, as data originates from a wide range of diverse sources, including archival materials, fieldwork observations, museum collections, and oral histories. These technologies enable the integration of such scattered information into a unified, interoperable network, thereby facilitating the systematic association of entities such as rituals, places, languages, and social practices, and allowing researchers to explore relationships and patterns that would otherwise remain fragmented or difficult to identify.

The establishment of a shared conceptual understanding through ontologies is particularly important, as key concepts such as kinship, identity, and ritual often carry nuanced and context-dependent meanings. Ontologies enable the formal definition of these concepts and the explicit specification of the relationships between them, thereby reducing ambiguity and supporting a common conceptual framework. This, in turn, facilitates clearer communication and more consistent data interpretation across researchers and research contexts. As an indicative example of this shared conceptual understanding, one may consider the properties attributed to marriage and the way in which these are modeled. Specifically, these properties may be represented either as direct relationships between the involved individuals or through an event-centric approach, as discussed in section 3.2.2.

The most obvious and direct advantage is related to resource discovery. The proposed method for representing anthropological research knowledge not only increases the amount of ethnographic information available on the Semantic Web, but also improves search and browsing capabilities for this type of data, while avoiding some of the problems that may arise from representing Indigenous Knowledge within Western-oriented repositories. Because the "institutionalised repositories hosting digital artefacts, follow Eurocentric categorisation systems and hardly consider different ways of organising representations of the world" (Mukumbira and Winschiers-Theophilus 2024). The technologies of the semantic web are particularly well-suited to enabling polyvocality, as they support the integration, representation, and interconnection of multiple perspectives within a shared, structured framework. Presumably, there are an infinite number of criteria for categorizing these resources, to provide accurate information to the end user of the system.

The second -and maybe the most important aspect- is of concern to the domain of comparative anthropology and relevant cultural studies. In this field, comparison comes with many challenges (Abramson and Gong 2020). Modeling information according to the principles of the semantic web unlocks the potential and the dynamics for comparative anthropology. Because, in the context of the semantic web, any information will not be available only as text, but also as meaning; and knowledge representation will have the cumulative character of a cohesive archive. This will allow researchers to set sophisticated criteria while studying certain aspects of human activities, since the process of matching these against others would be performed using computational methods.

Another advantage derives from the required formalism in knowledge representation for computational processes; further accuracy and granularity are encouraged in recorded descriptions and extracted conclusions. In addition, the capability of implementing reasoning processes upon the data also allows the validation of decisions in certain research cases.

An additional potential includes the arising possibilities regarding new ways of visualization of results, such as graphs. Data can be derived from pre-existing and codified knowledge which may now be presented in an alternative form. Admittedly "anthropologists make very little use [of] diagrams to model the social processes they study" (Carlson and Anderson 2007, 642), but this type of representation can be very helpful. Maybe the only area where formalisms and diagrams have been extensively used is kinship. Dwight Read is one of the scholars who has engaged with the subject most systematically and in depth. An indicative example is the study *Cognition, Algebra, and Culture in the Tongan Kinship Terminology* (Read and Bennardo 2007).

This approach, significantly considering the issue of semantic interoperability, enhances the interaction of anthropological research with other scientific domains. Formal models are essential within an interdisciplinary context due to both their potential for dealing with abstractiveness and their focus on details; when, at the same time, they can be adequately "realistic" so that they guide people from distinct backgrounds to turn to the research of common subjects (Van Der Leeuw 2004, 123).

"Knowledge is the anthropologist's stock-in-trade. […] anthropology is the description and representation of knowledge possessed by human groups" (Fischer 1994, 6). By representing this knowledge in semantic web terms, it can be recorded in a language-agnostic environment which is concept-centric, offering the potential to put the qualitative analysis at a more abstract level.

### *4.2 Requirements and challenges*

The representation of anthropological knowledge in the semantic web is not the type of task that can be performed by an individual; not even by small groups of experts. And it is not a task that involves computer and information scientists exclusively. It demands collaboration from both domains involved. Regarding the relationship between networks and the mathematical theory of graphs and anthropology there has been an ongoing discussion for decades. Since the '70s Wolfe has highlighted the flourishing of network thinking in anthropology (Wolfe 1978). Although his recommendations are not RDF-based, the basic rationale is the same. Since then, progress has not been significant, although two essential factors have been introduced. First, computers are now fully incorporated into everyday activities and knowledge about computers is widespread among almost all citizens of the Western world, thus even anthropologists are familiar with them. Second, and perhaps more essential, the pursuit of activities related to computer science is not always a low-level technical process but one that can be implemented at a higher level of abstraction. Even the codification might not require pre-existing programming knowledge; just logic. Thus, the epicenter of the process moves from technical matters to conceptualization.

For the codification of anthropological data, there is no need to develop ad hoc anthropological computational methods because this type of methods is not domain-dependent (Fischer 1994, 1–2). Consequently, there is no need for anthropologists to invent a new methodology; nevertheless, it is required that they develop computational thinking, according to Wing's viewpoint. In this case, computational thinking is not perceived as programming but as conceptualization; it is not about creating thinking machines, but about how humans think; and it is not a method for quantifying information, but a way of understanding how humans use their ideas for approaching and solving problems (Wing 2006). Anthropologists, however, must also develop a working understanding of the logic underlying computational representation. For example, in Figure 2, the relationship between two individuals who marry can be correctly modeled as *symmetric* when both its domain and range are defined as Person, meaning that any person can stand in this relation to any other person. If, however, the domain were restricted to men and the range to women, while the relation remained defined as symmetric, this would lead to a logical inconsistency in the model: since every woman would also be defined as a man. This example shows how inappropriate alignment between cultural categories and formal relational structures can generate unintended and misleading implications.

A second example, based on Figure 1, also highlights the need for anthropologists to be the ones to undertake the modeling. If there is a need to extend the specific graph, then a declaration could be added stating that *Malinowski studied the Trobriands*. This addition raises questions which can only be answered by anthropologists: Which is the relation for declaring "studied"? Where and by whom is it defined? Which is the relation for declaring that he conducted "field research"? And, again, where is it defined, and by whom? Are the Trobriands indeed the ones defined by HRAF in the following link: http://ehrafworldcultures.yale.edu/collection?owc=OL06 ? What are the properties -and who defines them- that relate the concepts in their culture? Correspondingly, what are the relationships that define kinship and should be specified in Figure 2 in order to represent a particular culture? As already mentioned, these are questions that cannot be answered by anyone other than anthropologists and the communities they work with.

Because anthropological research is fundamentally concerned with the articulation of conceptual and ontological frameworks that represent the complexities of social reality. However, this endeavor is inherently constrained by the nature of the empirical data upon which such frameworks are built. Ethnographic and anthropological datasets frequently contain sensitive personal information that cannot be ethically disclosed in full. As a result, a central methodological challenge emerges: how to construct and communicate robust conceptual models that remain grounded in empirical observation while safeguarding the privacy and dignity of research participants. This tension underscores the need for approaches that prioritize abstraction, anonymization, and the careful translation of lived experience into analytically meaningful categories.

In this context, the question shifts from whether empirical data can be shared to how elements of such data might be selectively and responsibly integrated within broader conceptual representations. Principles such as FAIR (Findable, Accessible, Interoperable, and Reusable) data stewardship offer a potential framework for addressing this issue, yet their application within anthropology remains underdeveloped. Many scholars exhibit skepticism toward data-sharing paradigms (Pels et al. 2018), often due to legitimate ethical concerns and disciplinary traditions that privilege contextual depth over standardization. Nevertheless, there is a growing need to explore methodologies that enable partial data inclusion while preserving ethical integrity. Advancing this discussion requires a critical engagement with both the epistemological foundations of anthropology and the evolving landscape of digital research practices.

The adoption of both FAIR and CARE (Collective Benefit, Authority to Control, Responsibility, Ethics) principles is essential for the responsible and effective management of cultural data. While FAIR principles emphasize technical aspects such as data standardization, interoperability, and machine-actionability, CARE principles foreground issues of governance, ethics, and the rights of communities, particularly indigenous and local stakeholders. Within this context, semantic web technologies provide a robust framework for implementing FAIR by enabling the formal representation of data through ontologies, linked data structures, and shared vocabularies, thereby enhancing discoverability and interoperability. At the same time, the use of linked data could support CARE principles by facilitating the inclusion of provenance metadata, access controls, and culturally sensitive annotations that respect community authority and ethical constraints. Thus, the integration of semantic web approaches with linked data practices offers a complementary pathway toward aligning technical excellence with ethical responsibility in data stewardship.

In summary, representing a culture requires the identification of its core concepts and the explicit modeling of their interrelationships within the appropriate cultural context, while adhering to FAIR and CARE principles in the management and use of the data from which these conclusions are derived. In OWL terminology, this entails the formal definition of classes and the relationships among them. Consequently, semantic technologies provide an effective framework for representing ethnographic information and anthropological knowledge.

## 5. Discussion

Certain sub-domains of artificial intelligence -such as machine learning, on which Generative AI models are based- are black boxes that represent and manage data in ways that are 'understandable' only by computers[6]. On the contrary, ontological knowledge representation is an assertion box whose representation is based on a human-centric approach and on codifying knowledge in ways that are readable both by humans and computers. But it is impossible to transfer expert knowledge without an established method. This is what Guarino means by saying that a knowledge base is not a repository of knowledge which was extracted from the experts' brains, like bequeathing knowledge, but it is the outcome of a modeling process (Guarino 1995, 625).

Knowledge representation may be at the core of social anthropology (Fischer 1994, 3), but it must be emphasized that ontologies are cultural constructions reflecting specific epistemological approaches and as such, they are not about the representation of an objective world nor about the objective representation of a relative world. Whatever the perceptions of anthropologists about the world may be, these perceptions could be the subject of attempted representations. Because the ontological representation to which we refer is no longer interested in a direct identification and characterization of reality via categories, but in a conceptualization of how reality is represented in any mind or rather through any mind (Kohne 2014, 87). In this case, the famous saying of Spinoza applies: "What Paul says about Peter tells us more about Paul than about Peter"; in non-textual representation the same principle applies: what we choose to describe expresses both ourselves and the others. This specified way of description has two particularly important advantages if compared to natural language: first, it is less open to misinterpretations, and, second, it creates more options for computational processing.

This study refers to approaches that lead to more formalistic methods which are closer to algorithmic processes. Formalism of this type is not intended to simplify meaning, but to represent meaning in a precise and unambiguous way. Thus, the expressiveness of a system is not dependent upon the level of its formalism but upon the level of the granularity it can support. In other words, any specifications or generalizations are handled by the degree of resolution allowed by any system; and not by its degree of formalism.

According to Van Der Leeuw, the value of formalistic modeling for social sciences is multiple. An advantage of these models is that they allow researchers to describe concisely a wide spectrum of relations with a level of precision which is usually not possible with the only alternative tool we have, i.e. natural language. In addition, the application of formalistic models is not restrained by specific domains. No less important in the context of social sciences is the fact that these models use a description language that is different from the one used to describe the phenomena of the modeling process. He also notes that certain types of formalistic models can precisely and concisely describe any changes occurring within a complex network of relations. Due to these characteristics, modeling is the ultimate fit-for-purpose technique for the formalization of dynamic theories concerning certain phenomena, which can later be compared against our own observations (Van Der Leeuw 2004).

Although the formalistic representation method presented in previous sections may be unfamiliar to the cultures studied by anthropologists, this is no hindrance to the recommended method of representing anthropological information in the context of the semantic web. Script is also strange to some cultures; yet, current anthropology uses it to describe them. Anthropological research is not conducted by means of the "others", but by means of the anthropologist's methods. Besides, the scientific research is "a strenuous and devoted attempt to force nature into the conceptual boxes supplied by professional education" (T. S. Kuhn 1970, 5). Therefore, the osmosis of information science theory and methodologies into social anthropology could trigger a Kuhnian paradigm shift, which introduces a new way of representing and managing anthropological knowledge.

[6] In response to the fact that many modern generative AI systems operate as "black boxes", Explainable AI was developed. Explainable AI refers to methods and techniques that make the decisions of machine learning models understandable to humans. It aims to reveal why a model produces a specific output instead of treating it as a black box. One common approach in Explainable AI is LIME (Local Interpretable Model-agnostic Explanations), which explains individual predictions by approximating the model locally with a simpler, interpretable model. For example, LIME can highlight which words in a text classification task most influenced a prediction, such as identifying "free" and "click" as key indicators of spam. LIME does not have access to the internal structure of the model. It treats the model as a black box and approximates it using simple, interpretable approaches.

Finally, it should be remarked that the more we dive into the transition to computational data processing, the more the issue of its transparency comes to the fore. According to Kemper and Kolkman, the more deeply algorithms are incorporated into the functioning of organizations, the more influential and less transparent they become. The creators of algorithms may make arbitrary choices which are opaque to interested stakeholders. In this way, algorithms reflect the bias of their creators and may impose certain preferences or positions against alternative options (Kemper and Kolkman 2018). This is the main reason transparency is a mandate for the creation and functioning of algorithms.

## 6. Conclusions

This article does not recommend a radical disengagement from the textual representation of the social sciences in general, nor, in particular, from anthropological knowledge. It tries to provoke discussion about whether script and text production are the only methods for recording and presenting the results of qualitative analysis in social science research. Evidence shows that alternative ways are also available. Here, the focus was on the semantic web and knowledge representation, demonstrating how they can contribute to representing, managing, and disseminating anthropological knowledge. The recommended approach does not deal with algorithms and ontological representations as autonomous, technical artifacts, but as complex socio-technical systems (Seaver 2018, 378).

Through the recommended modeling, the knowledge of domain experts may be used to identify context and represent it precisely and purposefully. The explicit representation of context allows for an analysis at multiple levels, namely micro, macro, and in-between, creating a highly effective system, especially when data exchange is involved, since the essential element of linked data and semantic technologies is the encouragement of cooperation (Oldman, Doerr, and Gradmann 2016). Aside from this aspect, the power of computers to bridge the gap between theory and data collection could amount to nothing if they are not assigned more complex tasks than the basic organization of our notes and the visualization of selected information in tables and diagrams (Kippen 1988, 319). It is up to anthropologists to decide how they wish to interact with computer science and information science, but their decisions will affect the ecosystem within which their knowledge could be enriched, documented, and disseminated among other scientific domains.

Having presented our arguments, the proposed ontological representation does what (ideally) anthropology itself does: it makes explicit the knowledge that remains implicit within individuals. In other words, knowledge representation -through ontologies and knowledge graph technologies- aligns with social anthropology in that it seeks to document, in a clear and systematic manner, what is already known within a group of individuals, and to use this documentation to draw conclusions by interpreting the knowledge that it has captured.

## References

Abrams, M.H. 2014. "What Is a Humanistic Criticism?" In *The Emperor Redressed: Critiquing Critical Theory*, University Alabama Press.

Abramson, Corey M., and Neil Gong. 2020. "Introduction: The Promise, Pitfalls, and Practicalities of Comparative Ethnography." In *Beyond the Case: The Logics and Practices of Comparative Ethnography*, Oxford University Press, 1–27. https://dx.doi.org/10.1093/oso/9780190608484.003.0001.

Antoniou, Grigoris, and Frank Van Harmelen. 2008. *A Semantic Web Primer*. 2ed ed. Cambridge, Mass: MIT Press.

Barley, Nigel. 2000. *The Innocent Anthropologist: Notes from a Mud Hut*. New York: Waveland Pr Inc.

Bjerre-Nielsen, Andreas, and Kristoffer Lind Glavind. 2022. "Ethnographic Data in the Age of Big Data: How to Compare and Combine." *Big Data & Society* 9(1): 1–6. doi:10.1177/20539517211069893.

Boellstorff, Tom. 2015. *Coming of Age in Second Life: An Anthropologist Explores the Virtually Human*. New edition. Princeton, NJ: Princeton University Press.

Carlson, Samuelle, and Ben Anderson. 2007. “What Are Data? The Many Kinds of Data and Their Implications for Data Re-Use.” *Journal of Computer-Mediated Communication* 12(2): 635–51. doi:10.1111/j.1083-6101.2007.00342.x.
Chansanam, Wirapong, Lan Thi Nguyen, Chunqiu Li, and Christopher Khoo Soo Guan. 2025. “Semantic Knowledge Graphs for Intercultural and Ethnographic Diversity in the Greater Mekong Subregion.” *Journal of Intercultural Communication* 25(3): 11–24. doi:10.36923/jicc.v25i3.1161.
Clifford, James, and George E Marcus, eds. 1986. *Writing Culture: The Poetics and Politics of Ethnography*. Berkeley: University of California Press.
Curran, John. 2013. “Big Data or ‘Big Ethnographic Data’? Positioning Big Data within the Ethnographic Space.” *Ethnographic Praxis in Industry Conference Proceedings* 2013(1): 62–73. doi:10.1111/j.1559-8918.2013.00006.x.
Doerr, Martin. 2003. “The CIDOC Conceptual Reference Module: An Ontological Approach to Semantic Interoperability of Metadata.” *AI Magazine* 24(3): 75. doi:10.1609/aimag.v24i3.1720.
Feldman, Joseph. 2017. “Big Data and Ethnology.” *Anthropology Today* 33(3): 1–2. doi:10.1111/1467-8322.12345.
Feyerabend, Paul. 2009. *Against Method*. 3rd ed. London: Verso.
Fischer, Michael D. 1994. *Applications in Computing for Social Anthropologists*. edition published in the Taylor&Francis e-Library, 2005. London: Routledge.
Forberg, Peter L. 2021. “From the Fringe to the Fore: An Algorithmic Ethnography of the Far-Right Conspiracy Theory Group QAnon.” *Journal of Contemporary Ethnography*: 08912416211040560. doi:10.1177/08912416211040560.
Foucault, Michel. 2007. *The Order of Things: An Archaeology of the Human Sciences*. London: Routledge.
Frické, Martin. 2019. “The Knowledge Pyramid: The DIKW Hierarchy.” *Knowledge Organization* 46(1): 33–46. doi:10.5771/0943-7444-2019-1-33.
Galeano, Eduardo. 2009. *Mirrors: Stories of Almost Everyone*. London: Portobello.
Geertz, Clifford. 1973. *The Interpretation of Cultures: Selected Essays*. Basic Books.
Gruber, Thomas R. 1995. “Toward Principles for the Design of Ontologies Used for Knowledge Sharing?” *International Journal of Human-Computer Studies* 43(5–6): 907–28. doi:10.1006/ijhc.1995.1081.
Guarino, Nicola. 1995. “Formal Ontology, Conceptual Analysis and Knowledge Representation.” *International Journal of Human-Computer Studies* 43(5–6): 625–40. doi:10.1006/ijhc.1995.1066.
Hine, Christine M. 2000. *Virtual Ethnography*. SAGE Publications Ltd.
Kemper, Jakko, and Daan Kolkman. 2018. “Transparent to Whom? No Algorithmic Accountability without a Critical Audience.” *Information, Communication & Society* 0(0): 1–16. doi:10.1080/1369118X.2018.1477967.
Kippen, James. 1988. “On the Uses of Computers in Anthropological Research.” *Current Anthropology* 29(2): 317–20.
Kohne, Jens. 2014. “Ontology, Its Origins and Its Meaning in Information Science.” In *Philosophy, Computing and Information Science*, eds. Ruth Hagengruber and Uwe Riss. London: Pickering & Chatto, 85–89.
Kuhn, Thomas S. 1970. *The Structure of Scientific Revolutions*. [2d ed., enl. Chicago: University of Chicago Press.
Kuhn, Tobias. 2014. “A Survey and Classification of Controlled Natural Languages.” *Computational Linguistics* 40(1): 121–70. doi:10.1162/COLI_a_00168.
Kuper, Adam. 2003. *Culture: The Anthropologists' Account*. 5. print. Cambridge, Mass.: Harvard Univ. Press.
Machado, Luís Miguel Oliveira, Renato Rocha Souza, and Maria da Graça Simões. 2019. “Semantic Web or Web of Data? A Diachronic Study (1999 to 2017) of the Publications of Tim Berners-Lee and the World Wide Web Consortium.” *Journal of the Association for Information Science and Technology* 70(7): 701–14. doi:10.1002/asi.24111.
Manganaro, Marc, ed. 1990. *Modernist Anthropology: From Fieldwork to Text*. Princeton, New York: Princeton University Press.
Marcus, George E. 1998. “The Once and Future Ethnographic Archive.” *History of the Human Sciences* 11(4): 49–63. doi:10.1177/095269519801100404.
Marcus, George E. 2007. “Ethnography Two Decades after Writing Culture: From the Experimental to the Baroque.” *Anthropological Quarterly* 80(4): 1127–45.
Margolis, Eric, and Stephen Laurence. 2011. “Concepts” eds. Edward N. Zalta and Uri Nodelman. *The Stanford Encyclopedia of Philosophy*. http://plato.stanford.edu/archives/spr2014/entries/concepts/ (May 2, 2025).
Mueller, Alain. 2016. “Beyond Ethnographic Scriptocentrism: Modelling Multi-Scalar Processes, Networks, and Relationships.” *Anthropological Theory* 16(1): 98–130. doi:10.1177/1463499615626621.
Mukumbira, Sebastian, and Heike Winschiers-Theophilus. 2024. “Implications of an Ecospatial Indigenous Perspective on Digital Information Organization and Access.” *International Journal on Digital Libraries* 25: 241–48. doi:10.1007/s00799-023-00353-6.
Munk, Anders Kristian, Mathieu Jacomy, and Asger Gehrt Knudsen. 2022. “The Thick Machine: Anthropological AI Between Explanation and Explication.” *Big Data & Society* 9(1). doi:10.1177/20539517211069891.
Oldman, Dominic, Martin Doerr, and Stefan Gradmann. 2016. “Zen and the Art of Linked Data.” In *A New Companion to Digital Humanities*, eds. Susan Schreibman, Ray Siemens, and John Unsworth. John Wiley & Sons, Ltd, 251–73. http://onlinelibrary.wiley.com/doi/10.1002/9781118680605.ch18/summary (June 18, 2025).

Pels, Peter, Igor Boog, J. Henrike Florusbosch, Zane Kripe, Tessa Minter, Metje Postma, Margaret Sleeboom-Faulkner, et al. 2018. “Data Management in Anthropology: The next Phase in Ethics Governance?” *Social Anthropology* 26(3): 391–413. doi:10.1111/1469-8676.12526.
Peponakis, Manolis, Sarantos Kapidakis, Martin Doerr, and Eirini Tountasaki. 2024. “From Calculations to Reasoning: History, Trends and the Potential of Computational Ethnography and Computational Social Anthropology.” *Social Science Computer Review* 42(1): 84–102. doi:10.1177/08944393231167692.
Peponakis, Manolis, Anna Mastora, Sarantos Kapidakis, and Martin Doerr. 2019. “Expressiveness and Machine Processability of Knowledge Organization Systems (KOS): An Analysis of Concepts and Relations.” *International Journal on Digital Libraries* 20(4): 433–52. doi:10.1007/s00799-019-00269-0.
Pinker, Steven. 2007. *The Language Instinct: How the Mind Creates Language*. third edition. New York: Harper Perennial Modern Classics.
Poli, Roberto, and Leo Obrst. 2010. “The Interplay Between Ontology as Categorial Analysis and Ontology as Technology.” In *Theory and Applications of Ontology: Computer Applications*, eds. Roberto Poli, Michael Healy, and Achilles Kameas. Springer Netherlands, 1–26. http://link.springer.com/chapter/10.1007/978-90-481-8847-5_1 (May 4, 2025).
Read, Dwight, and Giovanni Bennardo. 2007. “Cognition, Algebra, and Culture in the Tongan Kinship Terminology.” *Journal of Cognition and Culture* 7(1–2): 49–88. doi:10.1163/156853707X171810.
Ribes, David, and Geoffrey C. Bowker. 2009. “Between Meaning and Machine: Learning to Represent the Knowledge of Communities.” *Information and Organization* 19(4): 199–217. doi:10.1016/j.infoandorg.2009.04.001.
Rips, Lance J., Edward E. Smith, and Douglas L. Medin. 2013. “Concepts and Categories: Memory, Meaning, and Metaphysics.” In *The Oxford Handbook of Thinking and Reasoning*, eds. Keith J. Holyoak and Robert G. Morrison. Oxford: Oxford University Press, 177–209.
Seaver, Nick. 2017. “Algorithms as Culture: Some Tactics for the Ethnography of Algorithmic Systems.” *Big Data & Society* 4(2): 1–12. doi:10.1177/2053951717738104.
Seaver, Nick. 2018. “What Should an Anthropology of Algorithms Do?” *Cultural Anthropology* 33(3): 375–85. doi:10.14506/ca33.3.04.
Sowa, John F. 2009. *Knowledge Representation: Logical, Philosophical, and Computational Foundations*. Pacific Grove: Brooks/Cole.
Starn, Orin. 2022. “Anthropology and the Misery of Writing.” *American Anthropologist* 124(1): 187–97. doi:10.1111/aman.13677.
Van Der Leeuw, S. E. 2004. “Why Model?” *Cybernetics and Systems* 35(2–3): 117–28. doi:10.1080/01969720490426803.
W3C. 2004. *OWL Web Ontology Language Reference*. http://www.w3.org/TR/2004/REC-owl-ref-20040210/ (September 21, 2015).
W3C. 2014. *RDF 1.1 Concepts and Abstract Syntax: W3C Recommendation*. http://www.w3.org/TR/2014/REC-rdf11-concepts-20140225/ (May 30, 2025).
Wing, Jeannette M. 2006. “Computational Thinking.” *Communications of the ACM* 49(3): 33–35. doi:10.1145/1118178.1118215.
Wolfe, Alvin W. 1978. “The Rise of Network Thinking in Anthropology.” *Social Networks* 1(1): 53–64. doi:10.1016/0378-8733(78)90012-6.

## Appendix. Illustrative OWL code for Figure 2

Indicative OWL code corresponding to Figure 2 of the paper. The code provides a possible representation of the modeling underlying the figure. It is intentionally simplified and is intended to illustrate the potential of OWL for expressing social phenomena, rather than to constitute a formal or complete ontology.

```
<?xml version="1.0"?>
<rdf:RDF xmlns="http://OntologyExample#"
    xml:base="http://OntologyExample"
    xmlns:owl="http://www.w3.org/2002/07/owl#"
    xmlns:rdf="http://www.w3.org/1999/02/22-rdf-syntax-ns#"
    xmlns:xml="http://www.w3.org/XML/1998/namespace"
    xmlns:xsd="http://www.w3.org/2001/XMLSchema#"
    xmlns:rdfs="http://www.w3.org/2000/01/rdf-schema#">
  <owl:Ontology rdf:about="http://OntologyExample">
    <rdfs:comment>Indicative code corresponding to Figure 2 of the paper Toward Non-Textual Representation of Social
        Anthropology: Modeling Cultures as Knowledge Graphs. This highly simplified approach is intended to demonstrate the
        potential of OWL for modeling social phenomena, rather than to constitute a formal ontology.</rdfs:comment>
  </owl:Ontology>

  <!-- Classes  -->
  <owl:Class rdf:about="http://OntologyExample#Person"/>

  <owl:Class rdf:about="http://OntologyExample#Woman">
    <rdfs:subClassOf rdf:resource="http://OntologyExample#Person"/>
  </owl:Class>

  <owl:Class rdf:about="http://OntologyExample#Man">
    <rdfs:subClassOf rdf:resource="http://OntologyExample#Person"/>
  </owl:Class>

  <!-- Object Property -->
  <owl:ObjectProperty rdf:about="http://OntologyExample#is_married_to">
    <rdf:type rdf:resource="http://www.w3.org/2002/07/owl#SymmetricProperty"/>
    <rdfs:domain rdf:resource="http://OntologyExample#Person"/>
    <rdfs:range rdf:resource="http://OntologyExample#Person"/>
  </owl:ObjectProperty>

  <!-- Individuals  -->
  <owl:NamedIndividual rdf:about="http://OntologyExample#John">
    <rdf:type rdf:resource="http://OntologyExample#Man"/>
    <is_married_to rdf:resource="http://OntologyExample#Mary"/>
  </owl:NamedIndividual>

  <owl:NamedIndividual rdf:about="http://OntologyExample#Mary">
    <rdf:type rdf:resource="http://OntologyExample#Woman"/>
  </owl:NamedIndividual>

</rdf:RDF>
```